\documentclass[12pt]{iopart}

\usepackage{graphicx}
\usepackage{dcolumn}
\usepackage{bm}
\usepackage{amssymb}
\usepackage{epsf}
\usepackage{color}
\usepackage[utf8]{inputenc}

\begin{document}

\title[]{Charge density wave and superconductivity in 6R-TaS$_2$}

\author{Sudip Pal$^1$$^*\footnote{presently at 1 Physikalisches Institute, University of Stuttgart, Germany}$, Prakash Bahera$^1$, S. R. Sahu$^1$, Himanshu Srivastava$^2$, A. K. Srivastava$^{2,3}$, N. P. Lalla$^1$,Raman Sankar$^4$, A. Banerjee$^1$, and S. B. Roy$^1$}
\address{$^1$ UGC DAE Consortium for Scientific Research, University Campus, Khandwa Road, Indore-452001, India}
\address{$^2$ Synchrotrons Utilization Section, Raja Ramanna Centre for Advanced Technology, Indore-452013, India}
\address{$^3$ Homi Bhabha National Institute, Anushaktinagar, Mumbai 400094, Maharashtra, India}
\address{$^4$ Institute of Physics, Academia Sinica, Taipei 11529, Taiwan }
\ead{sudip.pal111@gmail.com,sindhunilbroy@gmail.com}

\begin{abstract}
The layered transition metal dichalcogenide compounds 1T-TaS$_2$ and 4H-TaS$_2$ are well known for their exotic properties, which include charge density wave, superconductivity, Mott transition, etc., and lately quantum spin liquid. Here, we report the magnetic, transport and transmission electron microscopy study of the charge density wave and superconductivity in 6R-TaS$_2$ which is a relatively less studied polymorph of this dichalcogenide TaS$_2$. Our high temperature electron microscopy reveals multiple charge density wave transitions between room temperature and 660K. Magnetization, and the electrical resistivity measurements in the temperature range of 2-400 K  reveal that 6R-TaS$_2$ undergoes a charge density wave transition around 305 K and is followed by a transition to a superconducting state around 3.5 K. The low temperature specific heat measurement exhibits anomaly associated with the superconducting transition around 2.4 K. The estimated Ginzburg Landau parameter suggests that this compound lies at the extreme limit of type-II superconductivity. 
\end{abstract}

\section{Introduction}
The coexistence {\color{blue}\cite{Zhu2011,Saleem2017}} or competition {\color{blue}\cite{Chang2012,Choi2020,Deguchi2012}} between the superconductivity and charge density wave (CDW) formation and their possible correlations with other interesting electronic and spin states, such as the Mott insulating state, antiferromagnetism, spin liquid etc., is a subject of intense investgations {\color{blue}\cite{Gabovich2002,Kurosaki2005}}. The high T$_c$ cuprate superconductors {\color{blue}\cite{Chang2012,Choi2020}}, chalocogenide based materials {\color{blue}\cite{Deguchi2012,Saleem2017}}, many organic compounds exemplify such interplay between different exotic electronic properties. Transition metal dichalcogenide (TMX$_2$, where TM is transition metals, and  X is S, Se, Te) is a class of materials, which hosts series of CDW transitions spread over a wide temperature range {\color{blue}\cite{Wilson1975,Rossnagel2011}}. In these compounds, the relation between the superconducting and CDW states is rather interesting. For example, in the classic CDW system 1T-TaS$_2$, suitable substitutions are found to suppress particularly the commensurate CDW state and simultaneously promote the superconductivity {\color{blue}\cite{Xu2010}}. In a few systems like 4Hb-TaS$_2$, 2H-NbSe$_2$ the two states are found to coexist {\color{blue}\cite{NbSe2,Freitas2016, Zheng2019,Leroux2015}}. Here we report coexistence of the charge density wave and superconductivity in 6R-TaS$_2$ a relatively less studied polymorph of TaS$_2$.

TMX$_2$s are of particular interest because of their layered quasi-two-dimensional crystal structure, where each layer consists of transition metals sandwiched between chalcogenides and the layers are bound by weak van der Waals force. The separation between the layers can be tuned {\color{blue}\cite{Morosan2007,Wang2018}}, and this in turn provides unique opportunities to study the physics in two dimension {\color{blue}\cite{Navarro2016}}. In addition, each TMX$_2$ layer can remain either in octahedral or trigonal prismatic coordination. Depending upon the stacking of the layers and size of a unit cell, different polytypes containing purely octahedral (1T type), purely trigonal prismatic (2H type), or mixed (4Hb, 6R type) coordination of metal ions in individual layers are possible {\color{blue}\cite{Wilson1975}}. 1T-TaS$_2$ has recently drawn considerable attention because it hosts a series of CDW transitions followed by a transition to Mott insulating state, a low-temperature quantum spin liquid state, glass-like resonating valance bond state, hidden electronic state, etc.,{\color{blue}\cite{Sipos2008,LawLee, JPCM, Stojchevska2014}}. Suitable substitution induces superconductivity in 1T-TaS$_2$ at very low temperatures. The 2H-TaS$_2$ and 4Hb-TaS$_2$ with their trigonal prismatic coordination show CDW and superconductivity in the pure compounds. In 2H-TaS$_2$, the CDW transition occurs at T= 80 K and superconductivity below T$_c$ = 0.7 K {\color{blue}\cite{Nagata1992}}. The 4Hb-TaS$_2$ consists of alternating layers of octahedral and trigonal prismatic coordination. It shows two CDW transition at T = 315 and 22 K and the chiral superconductivity below T$_c$= 2.7 K {\color{blue}\cite{DiSalvo1973, Nakashizu1984}}. Therefore, various structural factors like coordination, staking arrangement, and periodicity of TMX$_2$ layers appear to play a key role.

 There are very few studies available on the 6R-TaS$_2$ and this compound has remained largely unexplored. The charge density wave state near room temperature in 6R-TaS$_2$ has been discussed by Wilson  {\color{blue}\cite{Wilson1975}} and recently it is investigated in details {\color{blue}\cite{Amritroop2022}}. However, whether this compound is superconducting or not at low temperatures is still unclear.  In an initial study, no superconducting transition was reported in temperature dependence of resistivity {\color{blue}\cite{Thomson1975}}. In a subsequent study, appearance of superconducting transition was found below T = 2.3 K together with two CDW transitions nears room temperature and T = 80 K {\color{blue}\cite{Figueroa1992}}.  Liu et al observed superconducting transition only in Se doped 6R-TaS$_2$ {\color{blue}\cite{Liu2015}}. In a recent investigation, the 6R phase of TaS$_2$ was found to be superconducting, where the 6R phase has been obtained by heat treatment of 1T-TaS$_2$ {\color{blue}\cite{Amritroop2022}}. Therefore, the existence of superconducting transition in parent 6R-TaS$_2$ should be reinvestigated and if it exists then its superconducting properties need to be throughly studied, which is still missing in the literature.
 
Here, we have prepared single crystalline 6R-TaS$_2$ samples and throughly characterised phase purity by lab based and synchotron x-ray diffraction (XRD) of both single crystals and powder samples obtained from grinding of multiple single crystals. The crystal structure, and the lattice parameters were obtained from Rietveld refinement, which also confirms the single phase nature of the 6R-TaS$_2$. We combine the results of the structural, magnetic, and electrical transport measurements in a single crystal sample of 6R-TaS$_2$ over a wide range of temperatures from T = 0.1 K to 400 K. We investigated the CDW transition around T = 305 K, and our high temperature transmission electron microscopy (TEM) study for the first time reveals another CDW transition at higher temperatures. Importantly, here we have primarily focussed on the superconducting state occurring below  T = 3.5 K through detail resistivity measurements. We further characterized the superconducting state by low temperature specific heat measurement, which confirmed the bulk nature of the superconducting transition in this compound. By comparing the superconducting parameters of the other phases of TaS$_2$, we also have tried to highlight the possible role of the underlying crystal structure in controlling the physical properties of this famity of superconductors. 

\begin{figure}[t]
\centering
\includegraphics[scale=0.5]{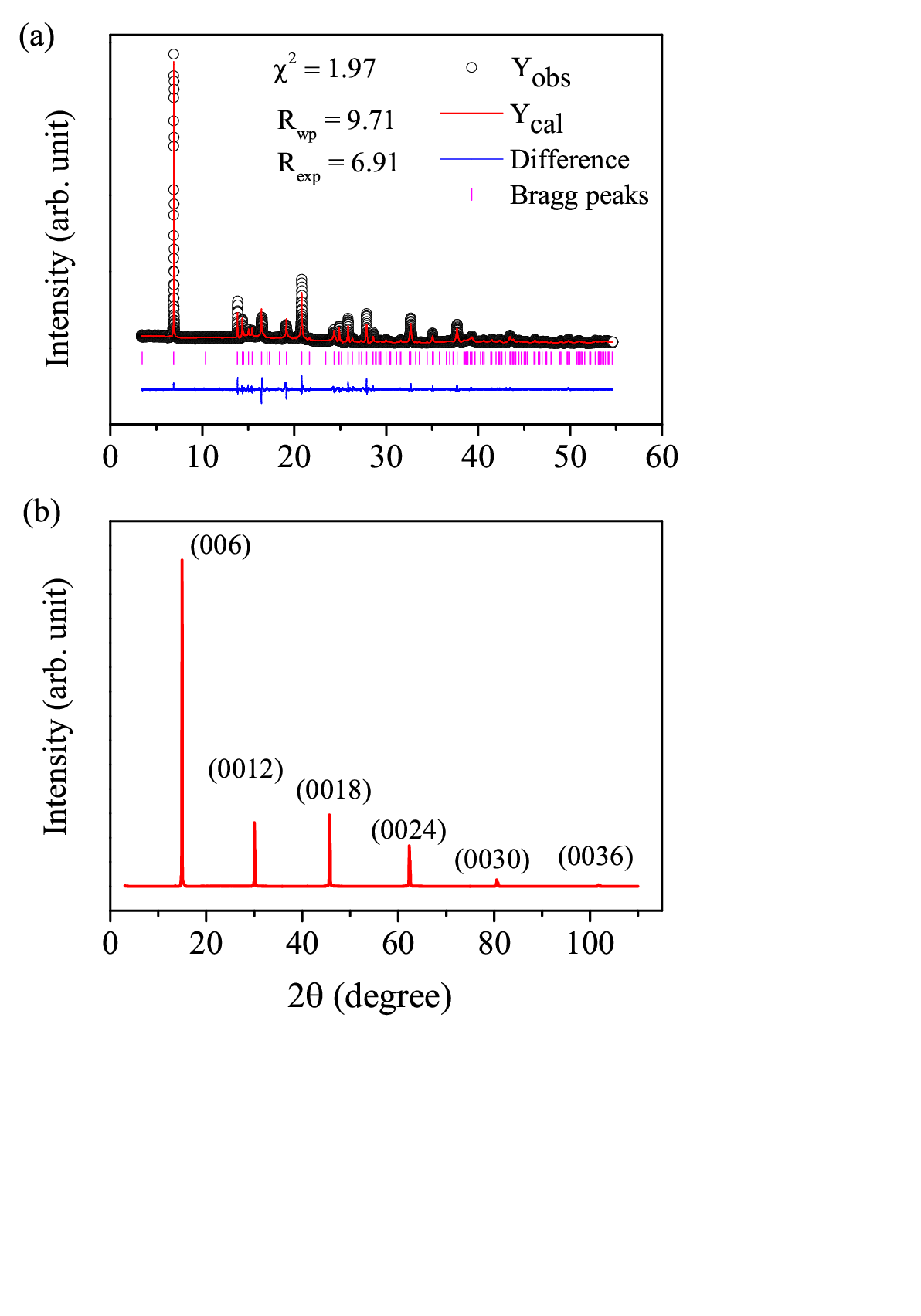}
\caption{(a) Synchotron x-ray diffraction pattern of crushed single crystals of 6R-TaS$_2$ recorded at wavelength $\lambda$ = 0.7178 \AA. (b) The XRD of single crystal flake using lab x-ray source of Cu K$_\alpha$.}
\end{figure}

\begin{figure*}[t]
\centering
\hspace{1cm}
\includegraphics[scale=0.64]{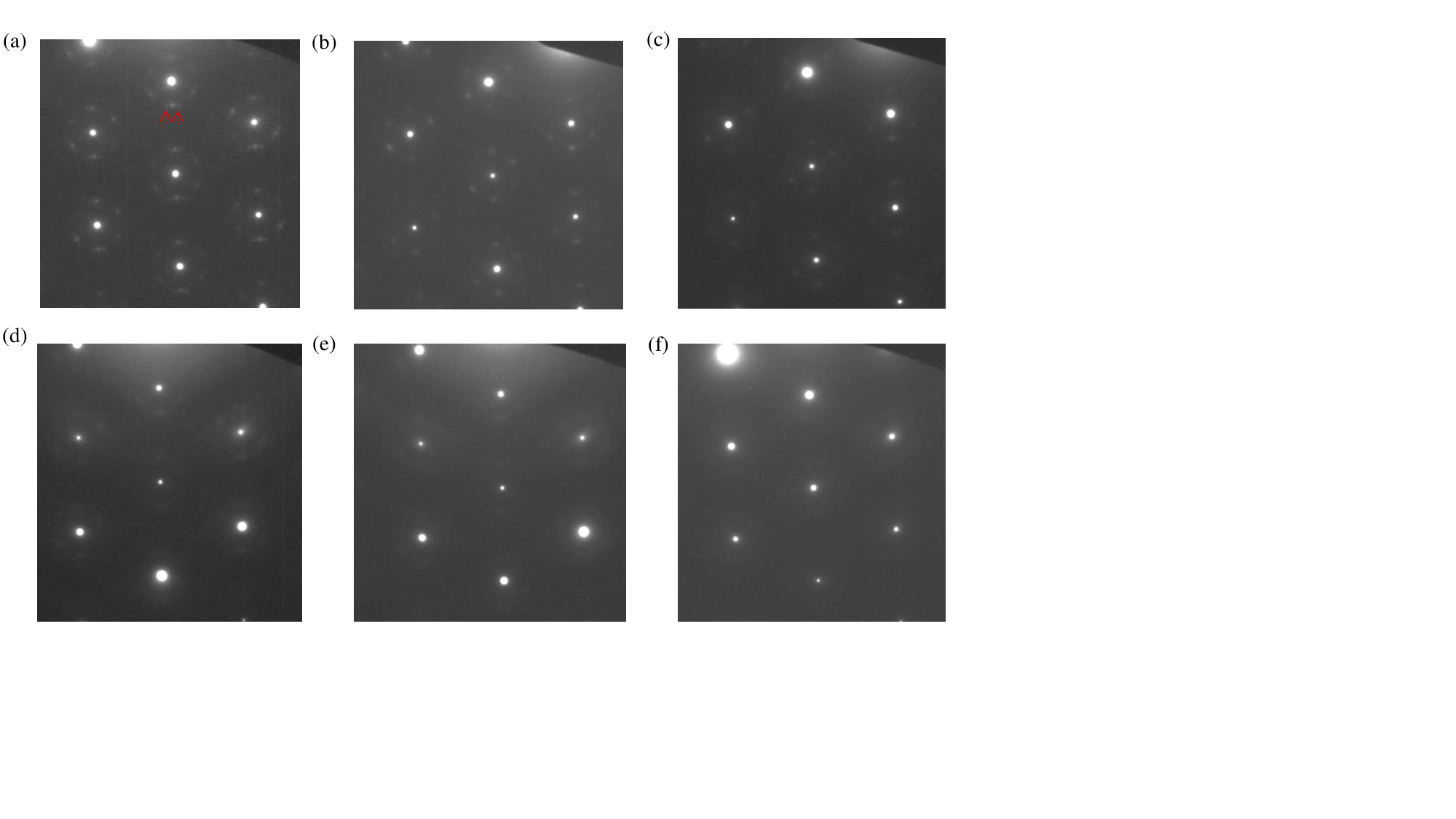}
\vspace{-3cm}
\caption{Temperature dependent TEM data at (a) T = 300 K (b) 310 K (c) 325 K (d) 570 K (e) 620 K and (f) 660 K. Two red arrows in (a) indicate the additional spots.}
\end{figure*}

\begin{figure}[t]
\centering
\includegraphics[scale=0.38]{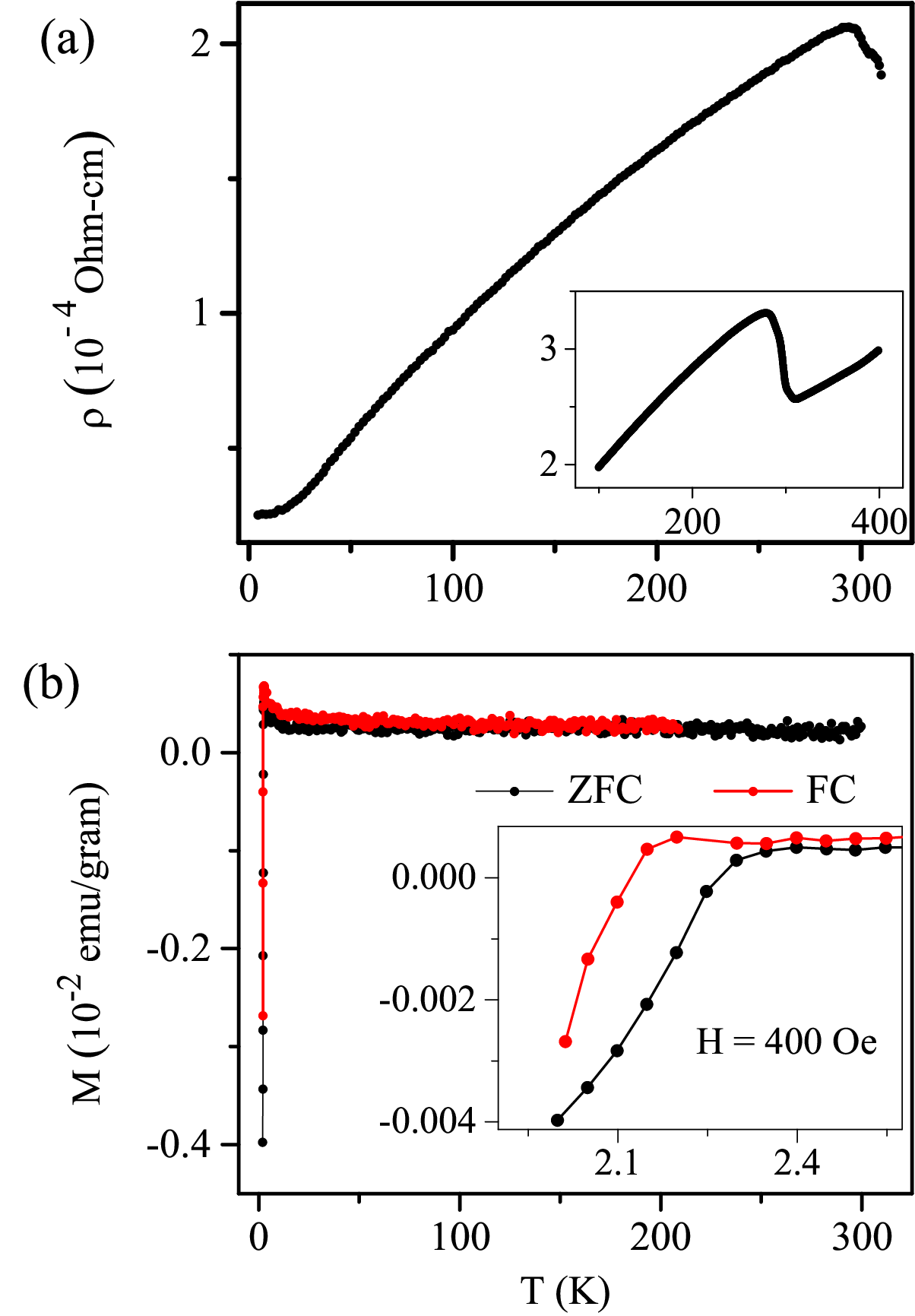}
\caption{(a) The main panel shows the temperature dependence of electrical resistivity of 6R-TaS$_2$ measured from 4.5 to 320 K in zero applied external field. The inset shows the expanded view of the temperature variation of resistivity (R-T) in the temperature region of T = 100 to 400 K highlighting the jump in resistivity. The inset figure has the same units as in the main panel. (b) Temperature dependence of magnetization ($M-T$) of 6R-TaS$_2$ recorded at the applied magnetic field of H= 400 Oe in ZFC and FC mode of measurement. Inset shows the expanded view of the $M-T$ near the superconducting transition. }
\end{figure}


\section{Experimental details}
The 6R-TaS$_2$ single crystals were grown from Chemical vapour Transport techniques (CVT) using Iodine as a Transport agent. Initially the polycrystalline powders of TaS$_2$ were prepared from the solid state synthesis process. The stoichiometric amount of high pure Ta slug (99.999\%) and Sulphur powder (99.999\%) were taken into the silica ampule and sealed at a high vacuum condition. The compound mixtures were heated at 600$^\circ$C and 800$^\circ$C for 24 hours and followed by intermediate grinding in an Argon filled glove box.  For the single crystal growth, the synthesized polycrystalline powders and 200 mg of iodine (I$_2$) were taken into the quartz ampule with the length of 40 cm and sealed under high vacuum (10$^{-3}$ Torr) condition. The quartz ampule was kept at horizontal two zone furnace. The charge end and growth end of the ampule were positioned at hot and cold zone of the furnace with the constant temperature of 850 and 750$^\circ$C for 200 hours. After completed the growth process, the temperature of the furnace was cooled down to room temperature at a rate of 2$^\circ$C/min. The good quality TaS$_2$ single crystals were harvested from the cold end of the ampule

The sample has been characterized by x-ray diffraction (XRD) studies in $\theta - 2\theta$ geometry using both lab-based and synchrotron x-ray sources. Synchrotron XRD study has been performed using beam line-12 in synchrotron radiation facility Indus-2 at Raja Ramanna Centre for Advanced Technology (RRCAT), Indore, India. Few pieces of single-crystalline flakes are carefully ground into fine powder for the synchrotron powder XRD which has been recorded at room temperature and the wavelength of $\lambda$ = 0.7178 \AA.  The lab XRD has been performed on a large single crystal flake taken from the same batch of samples using 18 kW rotating anode x-ray source producing Cu-$K_\alpha$ radiation {\color{blue}\cite{XRD}}. Transmission electron microscopy (TEM) has been performed on Philips CM200 using 200 keV electron energy. Magnetization (M) versus temperature (T) measurements of single-crystal samples have been performed using a 16-Tesla vibrating sample magnetometer (M/S Quantum Design, USA) in zero fields cooled (ZFC) and field cooled (FC) mode of measurements. In ZFC mode, initially, the sample is cooled down to the lowest temperature of measurement in the absence of an applied magnetic field. Then, the temperature dependence of M is recorded during the warming of the sample after the application of a constant magnetic field (H) at the lowest temperature. In the FC mode, the sample is first cooled to the lowest temperature of measurement in presence of the magnetic field and M has been recorded as a function of temperature while warming in the same field. Magnetization measurements have been corrected with a demagnetization factor of  N = 0.15 assuming rectangular cuboid shape of the sample {\color{blue}\cite{Demag2018}}. Electrical resistivity measurement within the temperature range of  T = 5 to 400 K has been performed using two different home-built setups which can used in the temperature range between 5-310 and 80-400 K respectively.  The resistivity measurement below 4 K has been performed in commercial 16 Tesla physical properties measurement system (PPMS) equipped with a dilution refrigerator (M/S Quantum Design, USA).  It may be noted here that in our electrical transport and magnetic measurements, we have used a few large single crystals (2-3 mm in length) from the same batch of samples. All the resistivity measurements have been performed using four-probe technique in van der pauw geometry {\color{blue}\cite{Pauw1958}}. Here we will present data obtained with two such flakes of 6R-TaS$_2$. The heat capacity for a sample of 5.4 mg was measured using a PPMS calorimeter (M/S Quantum Design, USA) equipped with Helium-3 refrigerator employing the relaxation method. The direction of the applied magnetic field is parallel to the a-b plane, like in the magnetization measurement.     

\section{Results and Discussion}
In Fig. {\color{blue}1(a)}, we have shown the XRD pattern of the powder sample of 6R-TaS$_2$ recorded at room temperature using a synchrotron x-ray source. The data have been analyzed by the Rietveld refinement method using Fullprof software package. All the reflections of powder XRD pattern can be fitted with R3m space group with lattice parameters, a=b= 3.34 $\AA$, c= 35.85 $\AA$ and $\alpha$ =$\beta$=90$^0$, $\gamma$ = 120$^0$. The parameters agree well with earlier reports {\color{blue}\cite{Jellinek1961}}. The unit cell consists of six S-Ta-S layers along the c-axis with alternating trigonal prismatic and octahedral coordination of Ta atoms. We have also shown the XRD pattern of the single crystalline flakes in Fig. {\color{blue}1(b)}, which is recorded at room temperature using a lab-based x-ray source. It shows intense peaks corresponding to (0,0,6) set of parallel planes. It indicates that the crystal is oriented along the c-axis.\\ 
To further characterize the structure, we have performed room temperature transmission electron microscopy (TEM) in imaging and diffraction modes. Electron transparent thin flakes for TEM investigations were obtained after repeated cleaving of TaS$_2$ crystal with the help of scotch tape.  Fig. {\color{blue}2(a)} shows a selected area electron diffraction (SAED) pattern taken along the c-axis at room temperature. The intense spots forming hexagonal patterns could be indexed by (hk0) type indices allowed in R3m space group. However, we note here that each of these spots is surrounded by another hexagonal set of less intense spots, nearly at one-forth distance. This reveals the formation of a superstructure as a result of the lattice modulation. The super-lattice spots have the typical arrangement as reported for 6R-TaS$_2$ phase {\color{blue}\cite{Wilson1975}}. Interestingly, each of these super-lattice spots is surrounded by two more spots, which are weakly visible, and have been marked by two arrows in Fig. {\color{blue}2(a)}. Now, looking at the earlier reports on other polytypes of TaS$_2$ and, this can be legitimately associated to the CDW formation, which is commensurate with the underlying lattice and occurs above room temperature. In Fig. {\color{blue}2(b)}-{\color{blue}2(f)}, we have shown TEM images at high temperatures. At T = 310 K, the hexagonal superlattice peaks around the main peaks are observed, but the two weak spots visible at room temperature are not clearly observed. As the temperature is increased further the intensity of the hexagonal superlattice peaks gradually reduces, but can still be observed even at T = 620 K. At T = 660 K, the superlattice peaks are no longer observed and the hexagonal pattern associated with the R3m space group is only visible. It indicates that the compound undergoes a CDW transition between T = 620 and 660 K. As temperature is reduced, it undergoes another CDW transition in between T = 300 and 310 K. It may be noted that the CDW transition temperature as observed here in 6R-TaS$_2$ is remarkably high as compared to other polytypes of TaS$_2$.

In the main panel of the Fig. {\color{blue}3(a)}, we have shown the temperature dependence of electrical resistivity, $\rho(T)$ of a single-crystalline flake of 6R-TaS$_2$ in the temperature range of T= 5 to 310 K.  The electrical contacts were made in the a-b plane. The sample is initially cooled down to T = 5 K and the data were recorded up to 310 K in the heating cycle and then cooled while measuring the resistivity in the cooling cycle. The temperature has been stabilized at each temperatures before recording the resistivity data. As we increase the temperature from T = 5 K, resistivity increases.  Above  T  = 296 K, resistivity starts decreasing rapidly till  T =  310 K. In addition, the $\rho(T)$ exhibits two distinct inflections, which indicates that there can be a possibility of two successive transitions in close vicinity. However, the origin of this anomalous two inflections is not clear. On the other hand, composition fluctuation of the single crystals may also be respopnsible for such behavior. Because in a separate crystal, where we have performed the high temperature resistivity measurement (inset of Fig. 3(a), discussed later), we did not observed such two features. In the subsequent cooling cycle, $\rho$(T) exhibits weak irreversibility producing small thermal hysteresis of width $\Delta$T $\approx$ 2 K. To get the idea of the high temperature behaviour we have measured $\rho$(T) in the temperature range between T = 100 to 400 K, which is the upper limit of our measurement system. The data has been shown in the lower inset of Fig. {\color{blue}3(a)}.  We note here that the resistivty values of the high temperature measurement (inset, Fig. 3a) is larger than the  resistivity shown in the main panel. It is because, the temperature dependence of resistivity in the temperature range of 100-400 K was performed on a different crystal than what is used to record the resistivity in the temperature range of 4to 310 K. As the temperature is reduced below T = 400 K, resistivity decreases which shows the metallic behavior of the sample. Around T = 305 K, resistivity shows a pronounced anomaly, where $\rho(T)$ suddenly increases and reduces at further lower temperature. The anomaly around T = 305 K is similar as observed in the $\rho(T)$ shown in the main panel of Fig. {\color{blue}3(a)}. This resistivity jump possibly arises to the CDW transition which is in consonance with the results of the TEM measurements at T = 300 and 310 K (see Figs. {\color{blue}2(a)} and {\color{blue}2(b)}). 

The weak thermal irreversibility as observed across the CDW transition is interesting. Recently, there is one report where similar irreversibility has been shown in the same compound {\color{blue}\cite{Amritroop2022}}. In general, thermal irreversibility in the cooling and heating cycles across a transition may arise due to first order nature of the transition. In addition, such a small irreversibility may appear due to experimental artifact, where the actual temperature of the sample can not follow the set temperature. So, to conform the genuineness of the thermal hysteresis, we have preformed resistivity measurements with multiple thermal cycles. We have included the details of measurement and data in the supplementary {\color{blue}\cite{Sup}}. We conclude that either there is no thermal hysteresis due to thermal lag in the experimental set up or it must be small. In this backdrop, it is important to note tha t the concept of CDW transition first addressed by Peierls-Frolich is based on weak electron-phonon coupling in purely one-dimensional system, which is now often applied to other low dimensional materials. Such transitions are supposed to be second order phase transition. However, in transition metal di-chalcogenides there are many systems where the CDW transition has been reported to be first order {\color{blue}\cite{Moncton1977,Stinitz1979,Adam2015}}. There, $\rho (T)$ exhibits prominent thermal hysteresis across the transition in the cooling and warming cycles. This requires a description beyond weak electron-phonon coupling. For example, the first-order CDW transition in TaSe$_2$ {\color{blue}\cite{Wilson1975}} can be semi-quantitatively understood via a microscopic theory proposed by McMillan {\color{blue}\cite{McMillan1977}}. This model invokes a short coherence length which leads to significant phonon softening. It is important to note that below the anomaly at T = 305 K, there is no further anomalous structure in the $\rho$(T) indicative of any kind of phase transition down to T = 5 K. However, the resistivity falls relatively fast below T = 50 K, and nearly saturates after T = 10 K. It may be noted that the variation of resistivity of 6R-TaS$_2$  with temperature is small and the absolute value is one order of magnitude less than the 1T-TaS$_2$ polytype and of the nearly same order to 2H-TaS$_2$ polytype {\color{blue}\cite{Fazekas1979}}. 

\begin{figure}[t]
\centering
\includegraphics[scale=0.36]{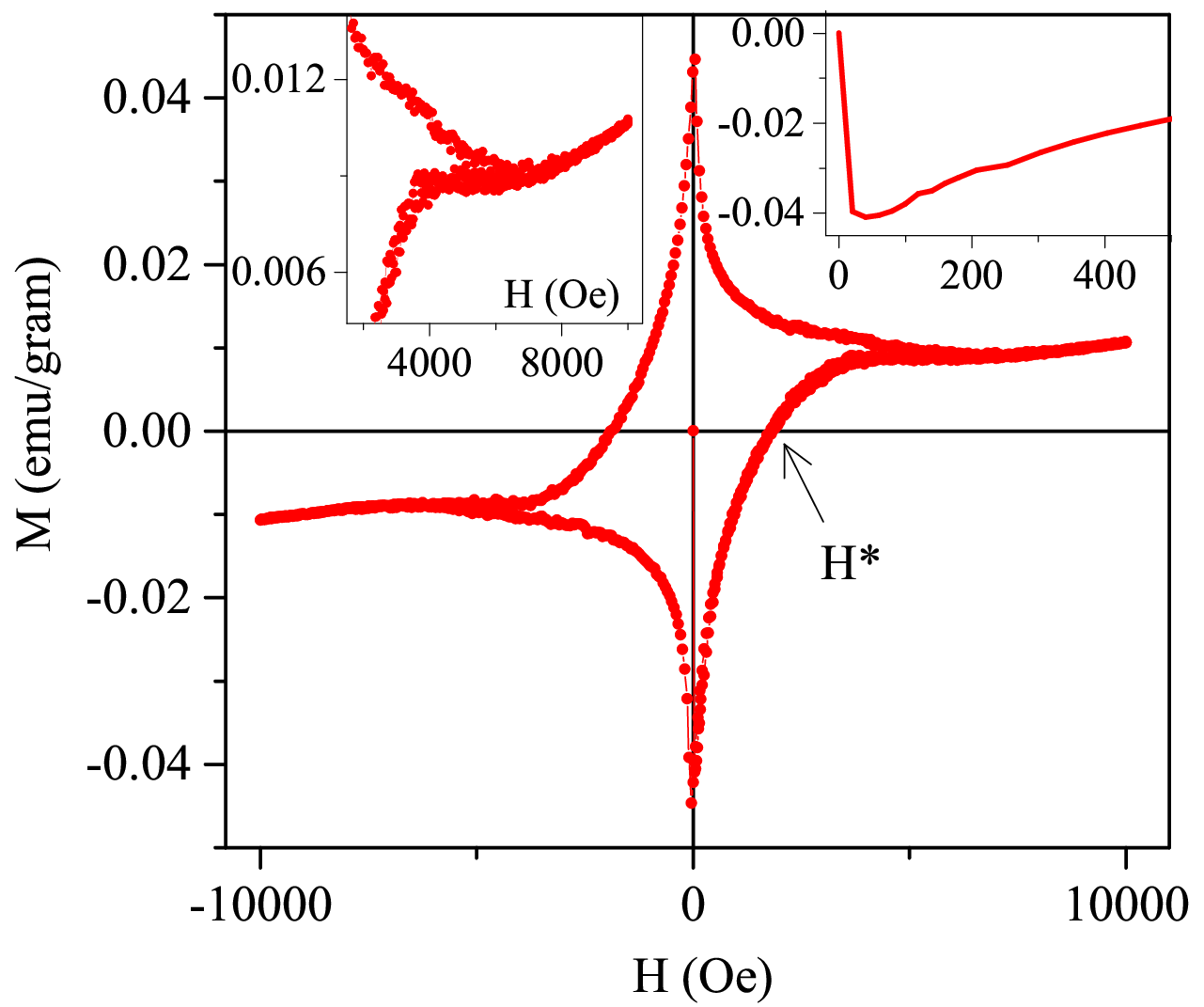}
\caption{Isothermal variation of magnetization (M) with external magnetic field (H) of 6R-TaS$_2$ at temperature T = 2 K in the zero field cooled state. The inset shows the expanded view of the virgin curve (first M-H cycle) near origin.}
\end{figure}
Fig. {\color{blue}3(b)} presents the temperature variation of M in 6R-TaS$_2$ measured with an applied magnetic field of H = 400 Oe in the ZFC and FC protocols. The field is applied parallel to the ab plane.  Unlike $\rho(T)$, the ZFC and FC magnetization curves do not show any appreciable anomaly around T$_{CDW}$ = 305 K. The magnetization curves are completely merged down to T = 50 K, and are nearly flat, which is consistent with the metal like the behavior of the sample. However, at further lower temperatures, M starts to gradually increase in both the ZFC and FC modes and there is a small opening (thermomagnetic irreversibility) between the ZFC and FC magnetization.  This reveals a non-trivial magnetic behavior below 50 K. Below around T = 2.4 K, magnetization exhibits diamagnetic response, which indicates a transition to a superconducting state. The transition temperature matches well with earlier reports where superconducting transition has been reported in the parent 6R-TaS$_2$ phase {\color{blue}\cite{Figueroa1992,Amritroop2022}}.  In the inset, we have shown the magnified view of the M-T response around the superconducting transition. In the ZFC mode, the sample tries to completely screen the external field giving rise to a diamagnetic response. In the FC mode, when the sample is cooled from a higher temperature than the T$_c$ in presence of a field, the sample tries to expel the external field below the transition temperature, which is the well-known Meissner effect. The transition temperature obtained from the ZFC curve is T$_c$= 2.4 K at H= 400 Oe. This is estimated from the downturn in the temperature dependence of magnetization. The thermomagnetic irreversibility observed between ZFC and FC magnetization curves in the superconducting state is reminiscent of magnetic flux line pinning within the sample, which in turn indicates that 6R-TaS$_2$ is a type II superconductor {\color{blue}\cite{John1993}}. The sample does not seem to achieve a complete Meissner state at 2 K. 

\begin{figure}[t]
\centering
\includegraphics[scale=0.32]{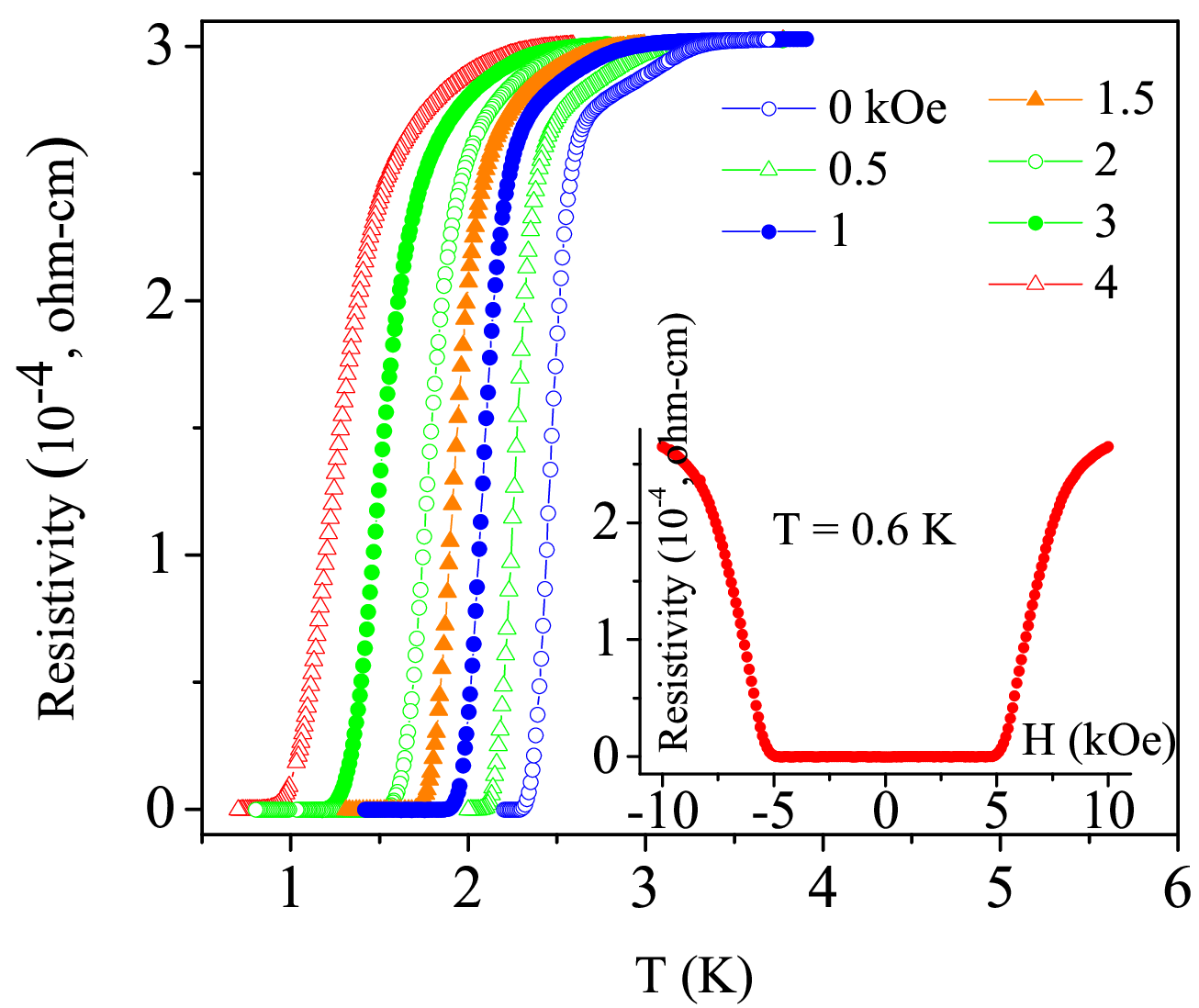}
\caption{Resistivity data of specimen 1 of 6R-TaS$_2$ showing superconducting transition. The main panel shows the effect of the field on the temperature variation of resistivity. Inset shows the isothermal resistivity data after zero-field cooling of the sample.}
\end{figure}
\begin{figure}[htb]
\centering
\includegraphics[scale=0.36]{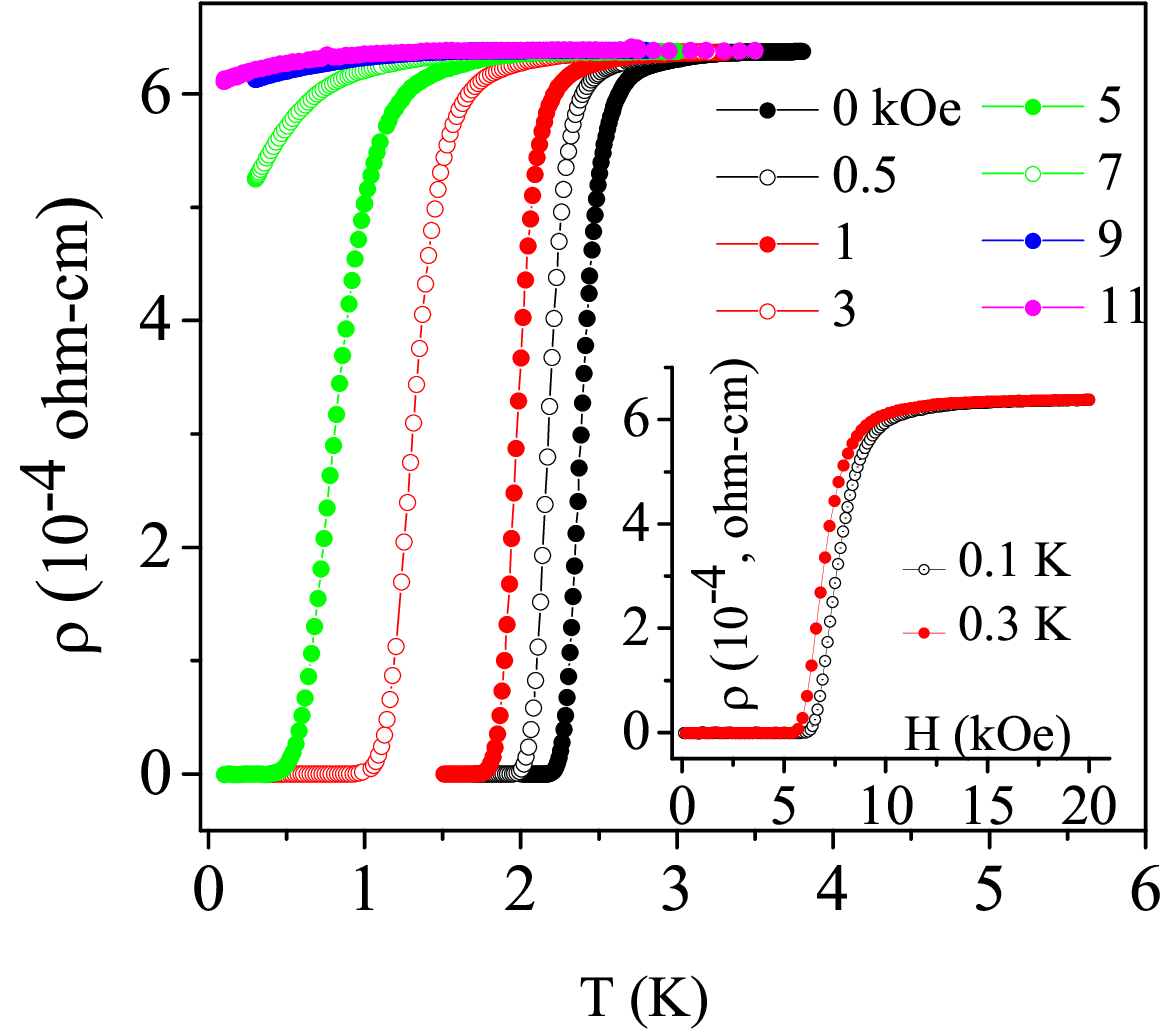}
\caption{Resistivity data of specimen 2  of 6R-TaS$_2$ showing the superconducting transition. The main panel shows the effect of field on the temperature variation of resistivity. Inset shows the isothermal resistivity data after zero field cooling of the sample.}
\end{figure}  
\begin{figure}[h]
\centering
\includegraphics[scale=0.38]{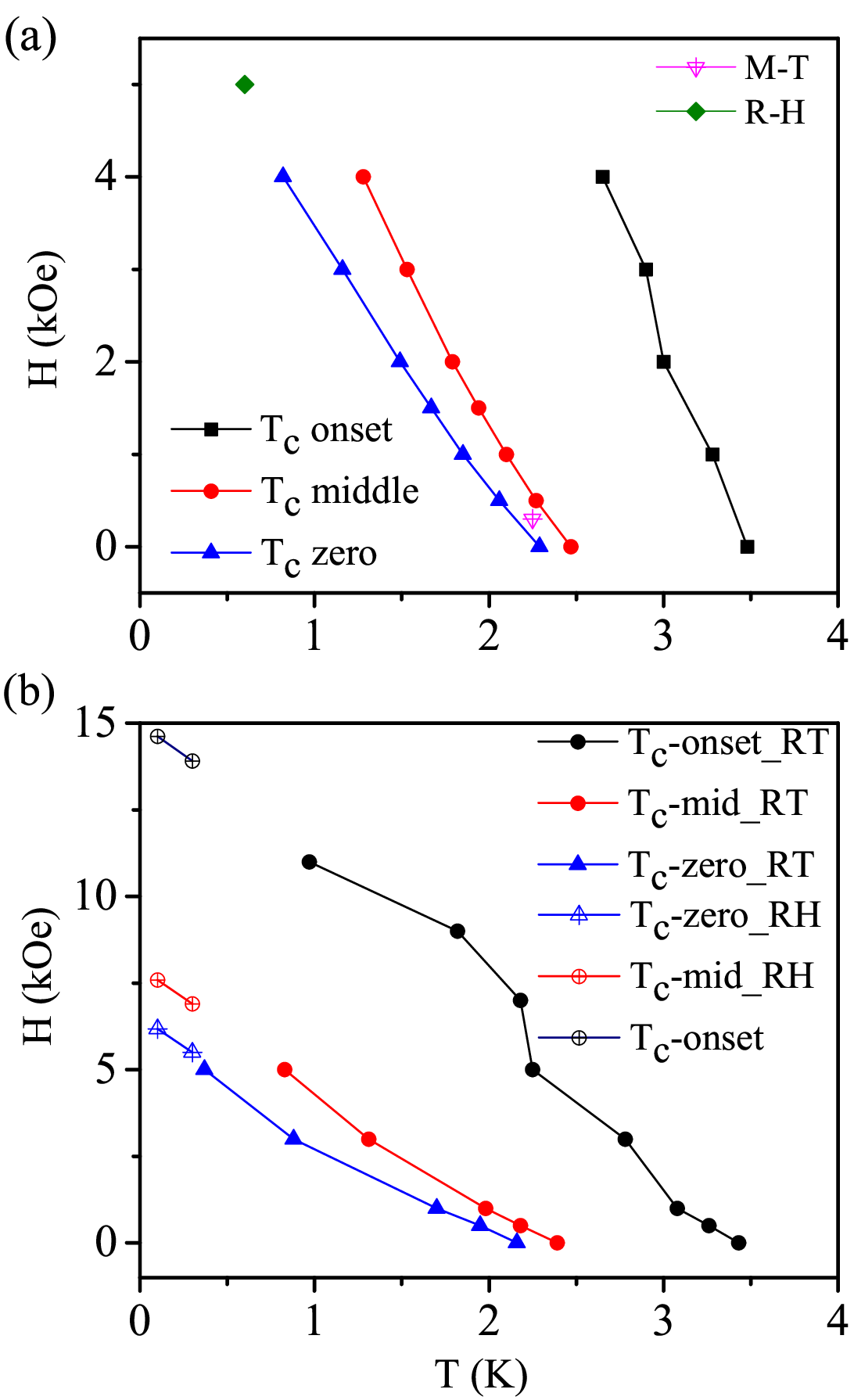}
\caption{Phase diagram of 6R-TaS$_2$ near superconducting region obtained from resistivity and magnetization measurements for two different flakes from the same batch, which have been obtained from the data shown in Fig. {\color{blue}5} and {\color{blue}6}. }
\end{figure}
\begin{figure}[h]
\centering
\includegraphics[scale=0.38]{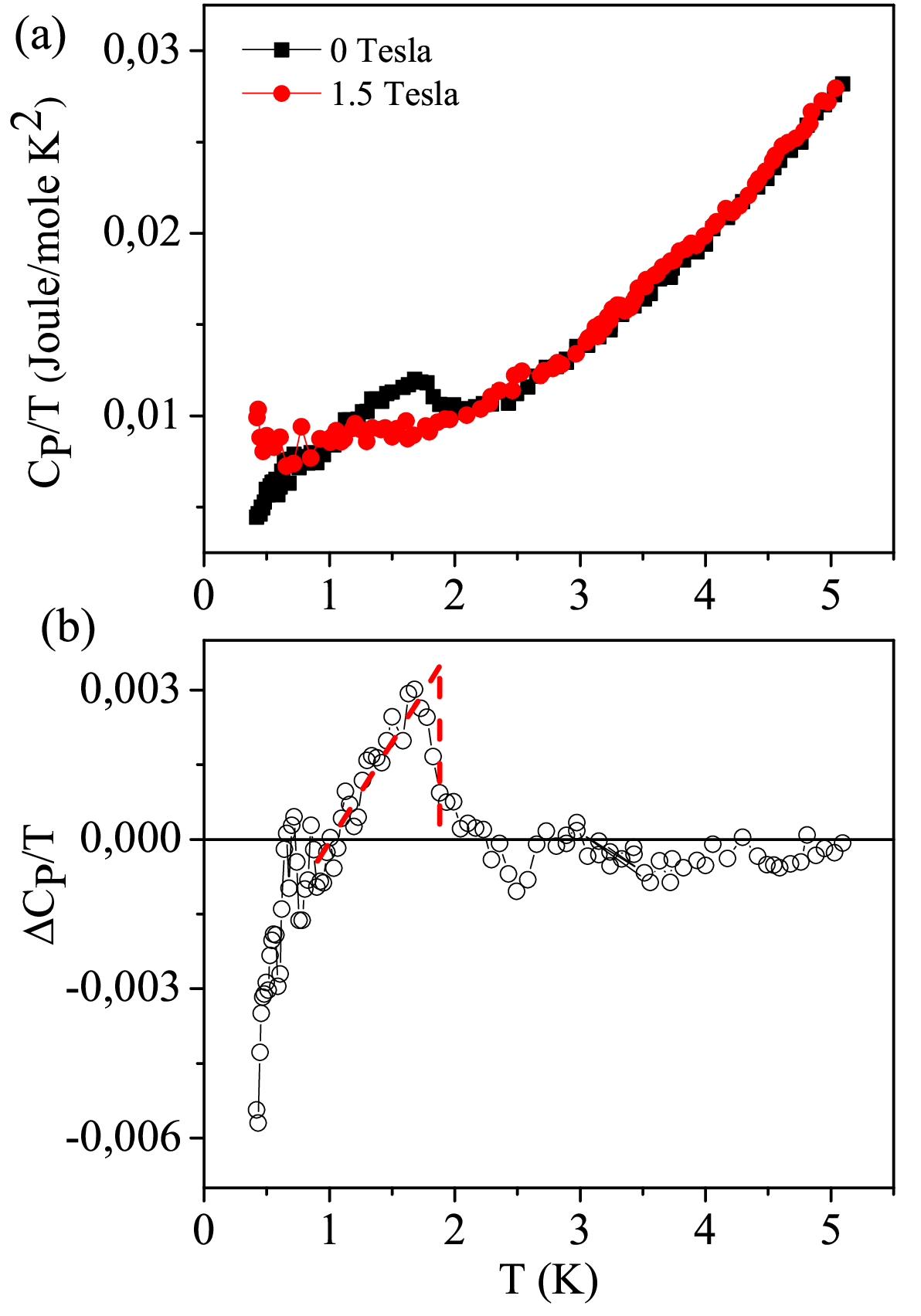}
\caption{(a) Temperature dependence of $C_P/T$ at H = 0 and 1.5 Tesla. (b) Temperature dependence of $\Delta C_P/T$, where $\Delta C_P$ = $C_P(H)-C_P(0)$. }
\end{figure}

The superconducting state is further investigated by measuring the isothermal M-H curve at T = 2 K after cooling the sample from room temperature in absence of an external field. We present the isothermal M-H curve in Fig. {\color{blue}4}. During measurement, the field is applied in the a-b plane. In the virgin magnetization cycle, M decreases with increasing field and then decreases, thus giving rise to a (negative) magnetization peak. The approximate value of the lower critical field H$_{c1}$ has been estimated from this peak, which is nearly H$_{c1}$ = 40 Oe, as shown in the inset of Fig. {\color{blue}4}. There are two interesting features in the M-H loop clearly showing the unusual nature of superconductivity in 6R-TaS$_2$. First, the field increasing and subsequent field decreasing branches merge at a positive value of magnetization, which indicates the presence of a strong paramagnetic component. The value of the field where the two branches merge is around H$_{irr}$ = 6 kOe, which is considerably larger than the field, H$^{*}$ = 1.8 kOe where magnetization becomes positive in the virgin cycle. Generally, in the case of the superconductors with paramagnetic contribution, H$_{c2}$ is estimated from the deviation from the linear normal state paramagnetic M-H curve {\color{blue}\cite{roy1996,roy2000}}. Here, the estimated value is around 8 kOe, which is consistent with the H$_{c2}$ obtained from the electrical resistivity measurements, as discussed later. This positive magnetization contribution in the superconducting state hints toward the nontrivial nature of the normal state of 6R-TaS$_2$ which is a CDW state. Secondly, in the virgin magnetization curve, after the first peak associated with H$_{c1}$, it shows a flat region followed by a shallow minimum at higher fields (see the inset of Fig. {\color{blue}4}). 

To check the genuineness of the occurrence of superconductivity in 6R-TaS$_2$ we have studied the temperature and magnetic field dependence of resistivity in two different single-crystal samples of 6R-TaS$_2$. In Figs. {\color{blue}5} and {\color{blue}6}, we present the temperature and magnetic field dependence of resistivity in 6R-TaS$_2$. Here, the current and voltage contacts were made on a-b plane and the magnetic field is applied perpendicular to a-b plane. Both the sample shows a transition into a superconducting state nearly below 3.5 K. The details of the temperature variation of H$_{c2}$ (H-T phase diagram)  for the sample 1 and 2 have been shown in Fig. {\color{blue}7} that is obtained from the $\rho$(T) and $\rho$(H) data shown in Figs. {\color{blue}5} and {\color{blue}6}, respectively. In R-T plots, the temperature where resistance deviates from the normal state resistivity is designated as T$_c$-onset. The temperature where the resistivity falls to 50\% and becomes zero are designated as T$_c$-middle and T$_c$-zero, respectively. We have also included the temperatures obtained from the R-H curves, shown in the inset. Superconducting transition temperature of both the sample is around, T$_c$ = 3.47 K. However, it is interesting to note that the value of T$_c$ found in resistivity measurement is significantly larger than the value obtained from the downturn in M-T measurement. In magnetization experiments, the diamagnetism might have started at a higher temperature but due to the presence of substantial paramagnetic contribution, the estimation of the exact temperature for the onset of diamagnetism is rather complicated. 

We observe that the temperature dependence of H$_{c2}$ is linear down to the lowest measured temperature. Using the relation, H$_{c2}$ = 1.84 $\times$ T$_c$, we obtain the Clogston-Chandrasekhar limit of upper critical field of 6 Tesla. On the other hand, using the Werthamer-Helfand-Hohenberg (WHH) relation $H_{c2} (T=0) = 0.69 \times (dH_c/dT)_{T}\times T_c$, the orbital limited upper critical field is around 1.1 Tesla, estimated for superconductor assuming singlet pairing and weak spin-orbit coupling in the system. This value is slightly lower than the upper critical field of 1.5 Tesla obtained from the linear extrapolation of data below T= 2 K in Fig. {\color{blue}7(b)}. This indicates pure s-wave like character of the superconductivity in 6R-TaS$_2$, similar to 4Hb-TaS$_2$ {\color{blue}\cite{Ribak}}. However, the slight shortfall of orbital limited value of $H_{c2}$ from the experimental value can be due to strong spin-orbit coupling {\color{blue}\cite{David2018}}. We have estimated  superconducting parameters using Gingzburg-Landau (GL) theory. The in plane (ab plane) coherence length ($\xi$) at T= 0 is found to be 148 $\AA$ which has been calculated by using the relation,  
\begin{eqnarray}
\xi = \sqrt{{\frac{\phi}{2\pi \times H_{c2}}}}
\end{eqnarray}
where, $\phi$ = 2.07 $\times$ 10$^{-15}$ Tesla-m$^2$. It is interesting to note that for 2H-TaS$_2$ and 4Hb-TaS$_2$, the value of in plane $\xi$ is around 360 and 185 $\AA$ respectively {\color{blue}\cite{Ribak,Yang2018}}. We also recall here that the CDW and superconducting transition temperatures monotonically increase from 2H-TaS$_2$, to 4Hb-TaS$_2$ and finally in 6R-TaS$_2$, which are different from each other in terms of the stacking of TaS$_2$ layers and the size of the unit cell. This monotonically decreasing value of coherence length along with increasing superconducting temperature points towards an intimate connection between superconductivity and the stacking of the trigonal prismatic TaS$_2$ layers. Much reduced coherence length indicates enhanced fluctuations near the phase boundary in the system. Besides, $\xi$, the other important length scale of the superconductors is the London penetration depth ($\lambda$). The value of $\lambda$ is $\lambda \approx$ 395 nm which has been estimated using the expression {\color{blue}\cite{Joshi2019}}:
\begin{equation} 
H_{c1} = \frac{\phi_0}{4\pi \lambda} (ln\frac{\lambda}{\xi} + 0.497)
\end{equation}

The estimated GL parameter is $\kappa = \lambda/\xi \approx $ 27, which indicates that 6R-TaS$_2$ belongs to the class of extreme type-II superconductors.

We have also recorded the temperature dependence of specific heat of 6R-TaS$_2$ in zero field and H = 1.5 Tesla down to T = 0.4 K. In zero-field measurements we observe an anomaly associated with the superconducting transition. According to the phase diagram in Fig. {\color{blue}7}, H = 1.5 Tesla is above the upper critical field within the measured temperature range, and thus sufficient to suppress the superconducting transition below T = 0.4 K. In Fig. {\color{blue}8(a)}, we have shown the temperature dependence of $C_P/T$ at H = 0 and 1.5 Tesla. The superconducting transition temperature $T_C$(0) = 2.4 K has been taken at which a deviation is observed from the normal state linearity in $C_P/T$ vs. $T^2$ plot. The value of $T_C$ is consistent with the magnetization measurement which is also performed with applied magnetic field parallel to the a-b plane. The normal state $C_P$ can be described by $C_P = \gamma T + \beta T^3$, where the first and second terms describe the electronic and lattice contributions to the specific heat respectively. The value of Sommerfeld coefficient of electronic heat capacity $\gamma$ is estimated to be about  7.6(7) mJ/moleK$^2$, and the Debye temperature $\theta_D$ = 115 K estimated from $\beta$ using $\theta_D = (1944Z_n/\beta)^{1/3}$, where $Z_n$ is the number of atoms per formula unit.  Fig. {\color{blue}8(b)} shows the temperature variation of $\Delta C_P/T$ = $C_P$ (T, H = 0) - $C_P$ (T, H)/T. The jump in the heat capacity, $\Delta C_P/T_C$ = 3.2 $\times$ 10$^{-3}$ J/moleK$^2$ across $T_C$ at zero field.

\section{Summary}
In summary, we have studied the transmission electron microscopy, electric transport, specific heat and magnetic properties of single-crystal samples of 6R-TaS$_2$. TEM at room temperature reveals a CDW state in this compound. High temperature TEM images reveal two CDW transition in between T = 300-310 K and 620-660 K, which is higher than other polytypes of TaS$_2$. Temperature dependent magnetization and resistivity data also show a CDW transition below 305 K which appears to be first-order in nature. The resistivity shows metal-like behavior in the CDW state.  Magnetization is nearly temperature independent within T= 300 to 50 K and shows a sharp rise at a further lower temperature. It is interesting to note that a superconducting state arises out of this enhanced paramagnetic state below 3.5 K.  This superconducting transition temperature is higher the superconducting transition temperatures in 2H-TaS$_2$ and 4Hb-TaS$_2$. The estimated GL parameter is around 27, which indicates the extreme type II nature of the superconductivity in 6R-TaS$_2$. 

\section{Acknowledgment}
We are thankful to Rohit Sharma for providing the single crystalline samples. We acknowledge Dr. Archana Sagdeo, RRCAT for the synchrotron XRD measurement, and Dr. V. Ganesan and M. P. Sarvanan for resistivity measurements in the dilution refrigerator.We acknowledge Dr. R. K. Kremer for the specific heat measurement. Sudip Pal and S. B. Roy acknowledge the Department of Atomic Energy (DAE), Government of India for the financial support in the form of Raja Ramanna Fellowship programme. 

\end{document}